\documentclass[aps,pre,twocolumn,superscriptaddress]{revtex4-1}

\usepackage{bm}
\usepackage{amsmath}
\usepackage{amssymb}
\usepackage{wasysym}
\usepackage{txfonts}
\usepackage{color}
\usepackage{graphics}
\usepackage{graphicx}
\usepackage{bbold}
\usepackage{xfrac}
\usepackage[colorlinks=true,linkcolor=blue,allcolors=blue]{hyperref}
\hypersetup{breaklinks=true}

\DeclareMathOperator{\E}{E}

\newcommand\Eq[1]{Eq.~(\ref{#1})}

\begin{document}

\title{Estimating the variance of Shannon entropy}

\author{Leonardo Ricci}
\email[]{leonardo.ricci@unitn.it}
\affiliation{Department of Physics, University of Trento, 38123 Trento, Italy}
\affiliation{CIMeC, Center for Mind/Brain Sciences, University of Trento, 38068, Rovereto, Italy.}
\author{Alessio Perinelli}
\affiliation{CIMeC, Center for Mind/Brain Sciences, University of Trento, 38068, Rovereto, Italy.}
\author{Michele Castelluzzo}
\affiliation{Department of Physics, University of Trento, 38123 Trento, Italy}

\date{\today}

\begin{abstract}
The statistical analysis of data stemming from dynamical systems, including, but not limited to, time series, routinely relies on the estimation of information theoretical quantities, most notably Shannon entropy. To this purpose, possibly the most widespread tool is provided by the so-called plug-in estimator, whose statistical properties in terms of bias and variance were investigated since the first decade after the publication of Shannon's seminal works. In the case of an underlying multinomial distribution, while the bias can be evaluated by knowing support and data set size, variance is far more elusive. The aim of the present work is to investigate, in the multinomial case, the statistical properties of an estimator of a parameter that describes the variance of the plug-in estimator of Shannon entropy. We then exactly determine the probability distributions that maximize that parameter. The results presented here allow to set upper limits to the uncertainty of entropy assessments under the hypothesis of memoryless underlying stochastic processes.
\end{abstract}

\maketitle
% Shannon entropy, stochastic processes, complexity, multinomial distribution

\section{Introduction}
\label{section:introduction}
The estimation of Shannon entropy~\cite{Shannon1948} associated to a probability distribution lies at the very core of information theory and is a fundamental tool in statistical analysis. Its appealing interpretation in terms of \emph{information content} makes Shannon entropy, along with metrics directly stemming from it, a crucial ingredient of many analytical techniques that are used to characterize complex systems. For example, mutual information, which is linked to the entropy difference between a joint distribution and the two related marginal distributions, is used to estimate the degree of correlation between two random variables~\cite{CoverThomas2006}. In the analysis of time series generated by complex systems, entropy measures such as permutation entropy~\cite{BandtPompe}, and approximate and sample entropy~\cite{Delgado_BonalMarshak2019} are widely used. Tools from information theory and based on entropy are common in neuroscience~\cite{Strong1998,Dimitrov2011,Sayood2018}, genomics~\cite{Vinga2014,Koonin2016,Chanda2020}, physiology~\cite{Burggren2005}, economics~\cite{Darbellay2000,Zapart2009}, climatology~\cite{Leung1990,Majda2010,Balasis2013}, geoscience~\cite{Wellmann2013,Perdigao2020}.

A key question is how reliably an estimator evaluates the actual entropy. The question becomes relevant especially in experimental situations when a single acquisition of limited size is available and thus only a single estimate can be assessed, for example in research fields that rely on historical time series or measurements of systems that cannot be controlled by the experimenter. While no general solution is known, the ways this issue is tackled depend on the---possibly unknown---physical mechanisms underlying the evolution of a system of interest. The issue of assessing an estimator's reliability is relevant also in wider contexts of information theory, such as Renyi entropies~\cite{Grassberger1988}, mutual information~\cite{Pardo1995}, and entropy rate~\cite{Lesne2009}.

The most basic case occurs when states are randomly and independently accessed, so that their number of visits follow a multinomial distribution. Despite its apparent simplicity, the issue of reliably estimating entropy in the multinomial case is far from being straightforward~\cite{Paninski}. Typically, Shannon entropy is estimated by relying on a recording of a system's discrete-time evolution: upon identifying the states of interest, the related occurrences are counted over an $N$-step evolution to yield a histogram of sample rates. Entropy is then evaluated on this empirical distribution by using the so-called plug-in estimator~\cite{AntosKontoyiannis} $\widehat{H}$, also referred to as maximum likelihood~\cite{Paninski} or naive~\cite{Strong1998} estimator. To assess the reliability of the plug-in estimator, crucial parameters are the support size $M$ and the sample size $N$. Early works by, among others, Basharin~\cite{Basharin}, Miller~\cite{Miller}, Harris~\cite{Harris} showed the consistency and the asymptotic normality of $\widehat{H}$.

In general, both the bias and the variance affecting $\widehat{H}$ depend on $N^{-1}$. However, while the bias is well-described by the so-called Miller-Madow correction~\cite{Miller,Basharin,Harris} equal to ${-(M-1)/(2N)}$ (see also Vinck et al.~\cite{Vinck2012} for more recent developments), there is no general rule to estimate the variance $\sigma^2_{\widehat{H}}$. More in detail, it is known that~\cite{Miller,Basharin,Harris}, upon defining the variance parameter ${\Lambda_0 = \sum_{i=1}^M s_i \left( H + \ln s_i \right)^2}$, where the $s_i$'s are the prior probabilities that define the multinomial distribution, and provided that ${\Lambda_0 \neq 0}$, one has ${\sigma^2_{\widehat{H}}\sim\Lambda_0/N}$. Unfortunately, whenever the multinomial distribution is unknown, so are $\Lambda_0$ and $\sigma^2_{\widehat{H}}$. Nevertheless, as shown by Antos and Kontoyiannis~\cite{AntosKontoyiannis}, $\sigma^2_{\widehat{H}}$ has an upper bound given by ${(\ln N)^2/N}$. In addition, Roulston~\cite{Roulston} derived, via standard error analysis, an estimate of $\sigma^2_{\widehat{H}}$ that has a widespread use (see for example~\cite{Kugiumtzis2013}).

In the present work we address the problem of estimating the variance parameter $\Lambda_0$ that describes $\sigma^2_{\widehat{H}}$ in the multinomial case. In the first part, we investigate the statistical properties of a \emph{straightforward} plug-in estimator $\widehat{\Lambda_0}$ for $\Lambda_0$. The estimator is shown to be itself consistent and asymptotically normal. Both its bias and variance are shown to be inversely proportional to the finite sample size $N$ making up the available data set.

In the second part, we derive an upper bound for the variance parameter $\Lambda_0$ that depends on the support size $M$. Provided that this support size is known, the upper bound can be used to set a limit to the degree of confidence on an estimate of $H$ via $\widehat{H}$ out of a sample histogram. The maximum variance parameter $\Lambda_{0,\,\text{max}}$ is shown to correspond to a distribution having a single, \emph{outlier} state, whereas the access probability to the remaining ones is uniformly distributed.

The paper is organized as follows. In Sec.~\ref{section:plugInLambdaZero}, upon summarizing the statistical properties of $\widehat{H}$, we introduce the plug-in estimator of the variance parameter $\Lambda_0$ and we derive its asymptotic distribution. Section~\ref{section:entropyVariance} concerns the assessment of the maximum variance parameter $\Lambda_{0,\,\text{max}}$ in the multinomial case and the related underlying distributions. Numerical simulations to test the reliability of the estimator of the variance parameter $\Lambda_0$ and to compare it with the one proposed by Roulston are discussed in Sec.~\ref{section:numericalExperiments}. Final remarks are presented in Sec.~\ref{section:conclusions}.

\section{Plug-in estimator of the variance parameter $\Lambda_0$}
\label{section:plugInLambdaZero}
\subsection{A summary on the plug-in entropy estimator $\widehat{H}$ and its statistical properties}
\label{subsection:plugInEntropy}
The system of interest consists of $M$ states and has a purely stochastic evolution: the transition probability from state $j$ to state $i$ is independent of $j$, where, here and henceforth, both $i$, $j$ run between 1 and $M$. Let $s_i$ be the probability with which the $i$-th state is visited. The Shannon entropy $H$ of the system is given by
\begin{equation}
	\label{eq:entropy}
	H = -\sum_{i=1}^{M} s_i \ln s_i \, .
\end{equation}
Throughout the paper it is assumed---via continuous extension---that ${x\ln^n x = 0}$ if ${x=0}$, for each ${n\in\mathbb{N}}$. However, unless otherwise specified, we suppose here that $M$ coincides with the size of the support of the distribution ${\{s_i\}}$, so that ${s_i > 0 , \,\forall i}$.

A way to estimate $H$ relies on the plug-in estimator $\widehat{H}$, as follows. (Henceforth, given a typically unknown quantity $a$, the symbol $\widehat{a}$ refers to an estimator of $a$ that, as a function of observed data, is a statistic affected by bias and fluctuations). We consider $N$ consecutive steps of the system's evolution and record the $M$-dimensional set ${\{\widehat{j}_i\}}$ of the number of visits of each one of the $M$ states. The set ${\{\widehat{j}_i\}}$ is distributed according to a multinomial distribution, and it holds that ${\sum_{i = 1}^M \widehat{j}_i = N}$. Upon computing, for each $i$, the observed rate $\widehat{p}_i$ as ${\widehat{p}_i = \widehat{j}_i/N}$, the plug-in estimator $\widehat{H}$ is defined as
\begin{equation}
	\label{eq:plugInEntropy}
	\widehat{H} \equiv -\sum_{i=1}^{M} \widehat{p}_i \ln \widehat{p}_i \, . \nonumber
\end{equation}

The plug-in estimator is affected by a bias term that is known as the Miller-Madow correction~\cite{Miller,Basharin,Harris} and is equal to ${-(M-1)/(2N)}$. With regard to the variance of the plug-in estimator, let first the variance parameter $\Lambda_0$ be defined as
\begin{equation}
	\label{eq:Lambda_0}
	\Lambda_0 \equiv \sum_{i=1}^M s_i \ln^2 s_i - \left( \sum_{i=1}^M s_i \ln s_i \right)^2 \, .
\end{equation}
Then, as shown by Basharin~\cite{Basharin} and Harris~\cite{Harris}, in the multinomial case and provided that ${\Lambda_0 > 0}$, the variance $\sigma^2_{\widehat{H}}$ is given by
\begin{equation}
	\label{eq:entropyVariance}
	\sigma_{\widehat{H}}^2 = \frac{\Lambda_0}{N} + O\left(\frac{1}{N^{3/2}}\right) \, . \nonumber
\end{equation}
Noteworthily, by writing $\Lambda_0$ in the form mentioned in Sec.~\ref{section:introduction}, namely ${\Lambda_0 = \sum_{i=1}^M s_i \left( H + \ln s_i \right)^2}$, it can be promptly shown that ${\Lambda_0 \geqslant 0}$, where the equality holds if and only if the probabilities $s_i$ are uniformly distributed.

As pointed out in the Introduction, estimating the variance of the plug-in estimator requires the knowledge of the parameter $\Lambda_0$, for which an estimator is proposed and analyzed in the following sections.

\subsection{Plug-in estimator $\widehat{\Lambda_0}$ of the variance parameter $\Lambda_0$}
\label{subsection:plugInLambdaZero}
The estimator of the variance parameter $\Lambda_0$ proposed in this paper is the plug-in estimator that is evaluated out of the rate histogram ${\{\widehat{p}_i\}}$ as
\begin{equation}
	\label{eq:sampleLambda0}
	\widehat{\Lambda_0} \left( \widehat{p}_1, \dots\, \widehat{p}_{M} \right) = \sum_{i=1}^{M} \left( \widehat{p}_i \ln^2 \widehat{p}_i \right) - \widehat{H}^2 \, .
\end{equation}
Because the quantity $\widehat{\Lambda_0}$ is a sample statistic, it is necessary to assess its reliability in terms of bias and variance. This issue is tackled by determining the asymptotic behavior of $\widehat{\Lambda_0}$ as ${N\to\infty}$. To this goal, it is suitable to define, for each $i$, ${\zeta_i \equiv \left( \widehat{p}_i - s_i \right) \sqrt{N}}$, so that the rates can be expressed as
\begin{equation}
	\label{eq:rateThroughZeta}
	\widehat{p}_i = s_i + \frac{\zeta_i}{\sqrt{N}} \, . \nonumber
\end{equation}
It follows:
\begin{eqnarray}
	\label{eq:sampleLambda0Zeta}
	\widehat{\Lambda_0} &=& \sum_{i=1}^{M} \left[ \left( s_i + \frac{\zeta_i}{\sqrt{N}} \right)\ln^2 \left( s_i + \frac{\zeta_i}{\sqrt{N}} \right) \right] \nonumber \\
		&-& \left[ \sum_{i=1}^{M} \left( s_i + \frac{\zeta_i}{\sqrt{N}} \right)\ln \left( s_i + \frac{\zeta_i}{\sqrt{N}} \right) \right]^2 \, . \nonumber
\end{eqnarray}
Expanding the previous expression in terms of order $N^{-n/2}$, with ${n\in\mathbb{N}}$, leads to
\begin{eqnarray}
	\label{eq:sampleLambda0TaylorSeries}
	\widehat{\Lambda_0} &=& \Lambda_0 + \frac{1}{\sqrt{N}} \sum_{i=1}^{M} \zeta_i \left[ \ell_i^2 + 2 (1 + H) \ell_i \right] \nonumber \\
		&+& \frac{1}{N} \left[ \sum_{i=1}^{M} \left( \frac{1 + H + \ell_i}{s_i} \zeta_i^2 \right) - \sum_{i=1}^{M} \sum_{j=1}^{M} \left( \zeta_i \zeta_j \ell_i \ell_j \right) \right] \nonumber \\
		&+& O \left( \frac{1}{N^{3/2}}\right) \, , \nonumber
\end{eqnarray}
where ${\ell_i\equiv\ln s_i}$ and the identity ${\sum_{i=1}^{M} \zeta_i = 0}$, which follows from the constraint ${\sum_{i=1}^M \widehat{p}_i = 1}$, was used.

Let ${G_{\delta\widehat{\Lambda_0}}(t)}$ be the moment generating function of the residual ${\delta\widehat{\Lambda_0} \equiv \widehat{\Lambda_0} - \Lambda_0}$, where $t$ is a real variable defined in a neighborhood of the origin. It holds that
\begin{equation}
	\label{eq:Lambda0_pgf}
	G_{\delta\widehat{\Lambda_0}}(t) = \E\left( e^{t \, \delta\widehat{\Lambda_0}} \right) \, . \nonumber
\end{equation}
The exponential form ${e^{t \, \delta\widehat{\Lambda_0}}}$ can be expressed as
\begin{eqnarray}
	\label{eq:exponentTaylorSeries}
	e^{t \, \delta\widehat{\Lambda_0}} &=& 1 + \frac{t}{\sqrt{N}} \sum_{i=1}^{M} \zeta_i \left[ \ell_i^2 + 2 (1 + H) \ell_i \right] \nonumber \\
		&+& \frac{t}{N} \left[ \sum_{i=1}^{M} \left( \frac{1 + H + \ell_i}{s_i} \zeta_i^2 \right)- \sum_{i=1}^{M} \sum_{j=1}^{M} \left( \zeta_i \zeta_j \ell_i \ell_j \right) \right] \nonumber \\
		&+& \frac{t^2}{2N} \sum_{i=1}^{M} \sum_{j=1}^{M} \zeta_i \zeta_j \left[ \ell_i^2 + 2 (1 + H) \ell_i \right] \left[ \ell_j^2 + 2 (1 + H) \ell_j \right] \nonumber \\
		&+& O \left( \frac{1}{N^{3/2}}\right) \, . \nonumber
\end{eqnarray}

Because the starting vector of the system's evolution is randomly chosen according to the distribution ${\{s_i\}}$, the expected value of each $\zeta_i$ is given by~\cite{Ricci2021PRE} ${\E\left(\zeta_i\right) = O\left( 1/N \right)}$, so that the term proportional to ${t/\sqrt{N}}$ is absorbed within the term ${O \left( N^{-3/2} \right)}$. In addition~\cite{Ricci2021PRE}, one has
${\E\left(\zeta_i \zeta_j\right) = \chi_{i,j} + O \left( N^{-1/2} \right)}$, where ${\chi_{i,j} = \delta_{i,j} s_i - s_i s_j}$ is the covariance matrix of rates being distributed according to a multinomial distribution ($\delta_{i,j}$ is the Kronecker delta). It follows:
\begin{eqnarray}
	\label{eq:exponentTaylorSeriesExpectedValue}
	G_{\delta\widehat{\Lambda_0}}(t) &=&
		\frac{t}{N} \left[ \sum_{i=1}^{M} \left( \frac{1 + H + \ell_i}{s_i} \chi_{i,i} \right) - \sum_{i=1}^{M} \sum_{j=1}^{M} \left( \chi_{i,j} \ell_i \ell_j \right) \right] \nonumber \\
		&+& \frac{t^2}{2N} \sum_{i=1}^{M} \sum_{j=1}^{M} \chi_{i,j} \left[ \ell_i^2 + 2 (1 + H) \ell_i \right] \left[ \ell_j^2 + 2 (1 + H) \ell_j \right] \nonumber \\
		&+& O \left( \frac{1}{N^{3/2}}\right) \, .
\end{eqnarray}
Let the coefficients of ${t/N}$ and ${t^2/(2N)}$ in \Eq{eq:exponentTaylorSeriesExpectedValue} be defined as $\gamma$ and $\Gamma$, respectively:
\begin{eqnarray}
	\label{eq:gamma_Gamma_definitions}
	\gamma &\equiv& \sum_{i=1}^{M} \left( \frac{1 + H + \ell_i}{s_i} \chi_{i,i} \right) - \sum_{i=1}^{M} \sum_{j=1}^{M} \left( \chi_{i,j} \ell_i \ell_j \right) \, , \nonumber \\
	\Gamma &\equiv& \sum_{i=1}^{M} \sum_{j=1}^{M} \chi_{i,j} \left[ \ell_i^2 + 2 (1 + H) \ell_i \right] \left[ \ell_j^2 + 2 (1 + H) \ell_j \right] \, . \nonumber
\end{eqnarray}
Equation~(\ref{eq:exponentTaylorSeriesExpectedValue}) can be then rewritten as:
\begin{equation}
	\label{eq:exponentTaylorSeriesExpectedValue_bis}
	G_{\delta\widehat{\Lambda_0}}(t) = \frac{t}{N} \gamma + \frac{t^2}{2N} \Gamma + O \left( \frac{1}{N^{3/2}}\right) \, .
\end{equation}
It is worth noting that, because $\chi$ is positive semi-definite, from the definition of $\Gamma$ one has ${\Gamma \geqslant 0}$.

As a consequence of the properties of the moment generating function, the coefficient of $t$ in this last equation corresponds to the expected value of $\delta\widehat{\Lambda_0}$, i.e. of ${\widehat{\Lambda_0} - \Lambda_0}$:
\begin{equation}
	\label{eq:deltaEstimator_populationMean}
	\mu_{\delta\widehat{\Lambda_0}} = \mu_{\widehat{\Lambda_0}} - \Lambda_0 = \frac{\gamma}{N} + O \left( \frac{1}{N^{3/2}}\right) \, .
\end{equation}
Also, because the square of this last term is of order $N^{-2}$, the coefficient of $t^2/2$ in \Eq{eq:exponentTaylorSeriesExpectedValue_bis} directly provides the variance of $\delta\widehat{\Lambda_0}$ and thus of $\widehat{\Lambda_0}$:
\begin{equation}
	\label{eq:estimator_populationVariance}
	\sigma_{\widehat{\Lambda_0}}^2 = \sigma_{\delta\widehat{\Lambda_0}}^2 = \frac{\Gamma}{N} + O \left( \frac{1}{N^{3/2}} \right) \, .
\end{equation}
Equations~(\ref{eq:deltaEstimator_populationMean}),~(\ref{eq:estimator_populationVariance}) provide the bias and the variance of the plug-in estimator $\widehat{\Lambda_0}$ of the variance parameter $\Lambda_0$, respectively.

Besides these results, \Eq{eq:exponentTaylorSeriesExpectedValue_bis} can be further exploited: starting from it, the moment generating function of the random variable ${w \equiv
\left( \delta\widehat{\Lambda_0} - \frac{\gamma}{N} \right) \sqrt{N}}$ is given by:
\begin{equation}
\label{eq:quasiStandardEntropy_pgf}
G_w(t) = \exp\left( \frac{t^2}{2} \Gamma \right) + O\left( \frac{1}{N^{1/2}} \right) \, , \nonumber
\end{equation}
where $t$ is a real variable defined in a neighborhood of the origin. This expression shows that, provided that ${\Gamma}$ is non-vanishing, $w$ is asymptotically normally distributed with zero mean and variance ${\Gamma}$. It is therefore possible to formulate a central limit theorem for the plug-in estimator $\widehat{\Lambda_0}$ of the variance parameter $\Lambda_0$:
\begin{equation}
\label{eq:CLT_entropy}
\widehat{\Lambda_0} \sim \mathcal{N} \left( \Lambda_0 + \frac{\gamma}{N}, \, \frac{\Gamma}{N} \right) \, \quad\quad\mbox{as }N\to\infty \, . \nonumber
\end{equation}
The main result of this section can be summarized as the fact that the plug-in estimator $\widehat{\Lambda_0}$ defined in \Eq{eq:sampleLambda0} is asymptotically normal and is a consistent estimator (both its population mean and variance depend on $N^{-1}$) of the variance parameter $\Lambda_0$.

\subsection{Practical considerations about $\widehat{\Lambda_0}$}
\label{subsection:plugInLambdaZero}
From a practical point of view, using the asymptotic expressions above requires the knowledge of the parameters $\gamma$ and $\Gamma$. The two parameters can be expressed upon straightforward algebra as
\begin{eqnarray}
	\gamma &=& M H + M - 1 - \Lambda_0 + \sum_{i=1}^{M} \ell_i \, , \label{eq:gamma} \\
	\Gamma &=& \mu_4^\prime - 4 \mu_3^\prime \left( H + 1 \right) + H^3 \left( 3 H + 4 \right) \nonumber \\
			&+& 2 \Lambda_0 \left( 3 H^2 + 6 H + 2 \right) - \Lambda_0^2 \, , \label{eq:Gamma}
\end{eqnarray}
where $\mu_3$, $\mu_4$ are the third and fourth moments of the distribution of the single-bin entropy ${-\ell_i = -\ln s_i}$, which are generally defined as
\begin{equation}
	\label{eq:moment_definition}
	\mu_n^\prime \equiv \sum_{i=1}^{M} s_i (-\ell_i)^n \, , \nonumber
\end{equation}
so that, for example, ${\mu_1^\prime = H}$, ${\mu_2^\prime = \Lambda_0 + H^2}$.

The presence, in the expressions above, of ${\mu_3^\prime}$, ${\mu_4^\prime}$, ${\sum_i \ell_i}$, makes the estimates of $\gamma$, $\Gamma$, and thus of $\widehat{\Lambda_0}$, a circular problem. In principle, provided that sufficiently large data samples are available, one could carry out a statistical analysis on subsamples of different size $N$, in order to ``experimentally assess'' the parameters $\gamma$, $\Gamma$. In practice however, and especially in the case of limited data samples, it is difficult to estimate the reliability of $\widehat{\Lambda_0}$ and thus of the variance of the estimated Shannon entropy $\widehat{H}$. In the next section this issue is circumvented by evaluating the maximum value of $\Lambda_0$ as a function of the number $M$ of the system's states.

\section{Maxima of the variance parameter $\Lambda_0$}
\label{section:entropyVariance}
The goal of this section is to find the maximum value of $\Lambda_0$ and the related probability distribution(s) ${\{s_i\}}$ for a given support size $M$.

Let $\mathbb{D}$ be the probability $M$-simplex defined as
\begin{equation}
	\label{eq:simplex}
	\mathbb{D} \equiv \left\lbrace \mathbf{r} \in \mathbb{R}^M \,\bigg\vert\, \sum_{i=1}^{M} r_i = 1 , \,r_i \geqslant 0 \, \forall i \right\rbrace \, . \nonumber
\end{equation}
Due to the settings ${x\ln^n x = 0}$ if ${x=0}$ and ${\forall n\in\mathbb{N}}$, which rely on continuous extensions of the respective functions, both $H$ and $\Lambda_0$ are continuous on the simplex $\mathbb{D}$. According to Weierstrass extreme value theorem~\cite{CourantHilbert} and the theorem of necessary condition for extreme values~\cite{AdamsEssex}, the extreme values of $\Lambda_0$ can only occur either on the boundary $\partial\mathbb{D}$ of $\mathbb{D}$ or in stationary points within $\mathbb{D}$. We first discuss the latter case.

\subsection{Stationary points in the interior of the probability simplex}
\label{subsection:withinSimplex}
Because the probabilities $s_i$ have to add up to one, only ${M-1}$ values are independent. We then consider $s_M$ as a function of the other $M-1$ probabilities:
\begin{equation}
	\label{eq:p0}
	s_M = 1 - \sum_{i=1}^{M-1} s_i \, .
\end{equation}

To find the stationary points of $\Lambda_0$, its partial derivative with respect to the independent $s_i$'s, with ${1 \leqslant i \leqslant M-1}$, is considered:
\begin{equation}
	\label{eq:dLambda0_dpi}
	\frac{\partial \Lambda_0}{\partial s_i} = \left(\ln s_i - \ln s_M \right) \left(\ln s_i + \ln s_M + 2 + 2H \right) \, . \nonumber
\end{equation}
A vanishing partial derivative requires fulfilling one of two possibilities, namely
\begin{eqnarray}
	s_i &=& p_0 \equiv s_M\, , \label{eq:solution_case1} \\
	s_i &=& q_0 \equiv \frac{1}{s_M}e^{- 2 - 2 H} \, . \label{eq:solution_case2}
\end{eqnarray}
The two possibilities are mutually-excluding: if it were ${p_0 = q_0}$, then it would follow, first, ${s_i = 1/M}$, ${\forall i \in [1,\,M]}$, so that ${H = \ln M}$, and, second, ${s_M = \frac{1}{s_M}e^{- 2 - 2 H}}$, so that ${H = 1 + \ln M}$. Therefore it must hold that ${p_0 \neq q_0}$.

Let $k$ be the number of $s_i$'s that satisfy the first possibility, \Eq{eq:solution_case1}. Because $s_M$ trivially satisfies this condition, one has ${1 \leqslant k \leqslant M}$. However, the case ${k=M}$ corresponds to a uniform distribution where ${s_i = 1/M}$, ${\forall i \in [1,\,M]}$. In this case, ${\Lambda_0 = 0}$ and, as shown by Harris~\cite{Harris}, the dominant variance term is of order $N^{-2}$. The uniform distribution thus yields the vanishing, minimum value of $\Lambda_0$, which also occurs in the $M$ vertices of the simplex, namely whenever a single $s_i$ is unitary while all the others are vanishing. Henceforth we therefore restrict the discussion to the range ${1 \leqslant k \leqslant M-1}$. The number of $s_i$'s that satisfy the second possibility, \Eq{eq:solution_case2}, is then ${M-k}$, which also varies between 1 and ${M-1}$.

Equations~(\ref{eq:entropy}),~(\ref{eq:p0}),~(\ref{eq:solution_case2}) can be respectively rewritten as follows:
\begin{eqnarray}
	-k p_0 \ln p_0 - (M-k)q_0 \ln q_0 &=& H \, , \label{eq:entropy_bis} \\
	k p_0 + (M-k) q_0 &=& 1 \, , \label{eq:p0_bis} \\
	\ln p_0 + \ln q_0 + 2 + 2 H &=& 0 \, . \label{eq:solution_case2_bis}
\end{eqnarray}
Equations~(\ref{eq:entropy_bis}),~(\ref{eq:p0_bis}),~(\ref{eq:solution_case2_bis}) make up a system of three equations in the three variables $H$, $p_0$, $q_0$, while the integer number $k$ is a fixed parameter. Inserting the expression for $H$ given by \Eq{eq:entropy_bis} into \Eq{eq:solution_case2_bis} and replacing $q_0$ with ${\frac{1 - k p_0}{M-k}}$, as given by \Eq{eq:p0_bis}, yields:
\begin{equation}
	\label{eq:condition}
	\left(1 - 2 k p_0 \right) \ln \frac{1 - k p_0}{k p_0} = 2 + \left(1 - 2 k p_0 \right) \ln \frac{M-k}{k} \, .
\end{equation}
Crucially, because of \Eq{eq:p0_bis} and the fact that we are considering the interior of $\mathbb{D}$, it holds that ${0 < k p_0 < 1}$ and ${0 < (M-k) q_0 < 1}$.

\begin{figure}[h]
	\centering
	\includegraphics[scale=1.0]{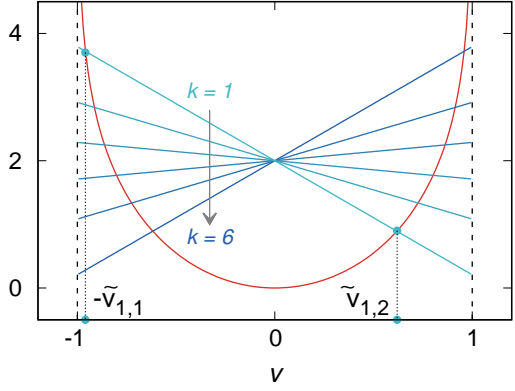}
	\caption{Graphical representation of \Eq{eq:condition_v}. The red, solid line corresponds to the graph of the function $f(v)$ defined in \Eq{eq:f_v_definition}. The straight-lines correspond to the right-hand term of \Eq{eq:condition_v} in the case ${M = 7}$ and for different values of the parameter $k$. The blue dots mark the intersection of $f(v)$ with the straight-line in the case ${k=1}$, as well as the two abscissae ${-\tilde{v}_{k,1}}$, ${\tilde{v}_{k,2}}$.}
	\label{fig:condition_v}
\end{figure}

It is useful to introduce the auxiliary variable $v$ defined as
\begin{equation}
	\label{eq:v_definition}
	v \equiv 2 k p_0 - 1 \, . \nonumber
\end{equation}
It follows:
\begin{eqnarray}
	\vert v \vert &<& 1 \, , \label{eq:v_range} \nonumber \\
	p_0 &=& \frac{1 + v}{2k} \, , \label{eq:p0_from_v} \\
	q_0 &=& \frac{1 - v}{2(M - k)} \, . \label{eq:q0_from_v}
\end{eqnarray}
By also defining the function $f(v)$ as
\begin{equation}
	\label{eq:f_v_definition}
	f(v) \equiv v \ln \frac{1+v}{1-v} \, ,
\end{equation}
\Eq{eq:condition} can be rewritten as
\begin{equation}
	\label{eq:condition_v}
	f(v) = 2 - v \ln \frac{M-k}{k} \, .
\end{equation}
The graph of $f(v)$ corresponds to the continuous, red line shown in Fig.~\ref{fig:condition_v}.
It is straightforward to show that ${f(v)}$ satisfies the following properties: ${f(-v)=f(v)}$; ${f(v) \geqslant 0}$; ${f(v) \to + \infty}$ as ${v \to \pm 1}$. In addition, because the derivative ${f^\prime(v)}$ is given by
\begin{equation}
	\label{eq:df_dv}
	f^\prime(v) = \frac{1}{v} \left[ f(v) + \frac{2v^2}{1-v^2} \right] \, , \nonumber
\end{equation}
the function ${f(v)}$ is monotonically increasing (decreasing) if ${v > 0}$ (${v < 0}$) and thus has a single minimum equal to 0 in ${v=0}$. As it is also shown in Fig.~\ref{fig:condition_v}, the right-hand term of \Eq{eq:condition_v} is instead a straight-line going through the point ${(0,2)}$ and having a slope ${-\ln \frac{M-k}{k}}$.

For any $k$, the roots of the transcendental \Eq{eq:condition_v} provide, via Eqs.~(\ref{eq:p0_from_v}),~(\ref{eq:q0_from_v}), stationary points for ${\Lambda_0 = \Lambda_0(p_0,p_1,\dots,p_{M-1})}$. Let $\tilde{v}_k$ be a root of \Eq{eq:condition_v} for a given $k$. According to \Eq{eq:Lambda_0} and by using \Eq{eq:p0_from_v},~(\ref{eq:q0_from_v}), the variance parameter $\Lambda_0$ can be promptly rewritten as
\begin{equation}
	\label{eq:Lambda_0_new}
	\Lambda_0 = \frac{1}{\tilde{v}_k^2} - 1 \, .
\end{equation}
It then follows that the maximum value of $\Lambda_0$ occurs for a stationary point characterized by a value $k$ that yields the minimum $|\tilde{v}_k|$.

To assess this $k$ value and the corresponding minimum $|\tilde{v}_k|$ we start observing that for each $k$ there are two roots of \Eq{eq:condition_v}, one negative and one positive, which are henceforth referred to as $-\tilde{v}_{k,1}$, $\tilde{v}_{k,2}$, respectively, with ${\tilde{v}_{k,1} > 0}$, ${\tilde{v}_{k,2} > 0}$. From \Eq{eq:condition_v} it holds that ${\tilde{v}_{M-k,2} = \tilde{v}_{k,1}}$. It is therefore sufficient to study the cases ${k \leqslant M/2}$ for which the factor ${\ln \frac{M-k}{k}}$ is non-negative.

Inserting ${-\tilde{v}_{k,1}}$, ${\tilde{v}_{k,2}}$ in \Eq{eq:condition_v} and subtracting the results yields
\begin{equation}
	\label{eq:inequality_1}
	f\left( -\tilde{v}_{k,1} \right) - f\left( \tilde{v}_{k,2} \right) = \left( \tilde{v}_{k,1} + \tilde{v}_{k,2} \right) \ln \frac{M-k}{k} \geqslant 0 \, . \nonumber
\end{equation}
Consequently, ${f\left(\tilde{v}_{k,1} \right) = f\left( -\tilde{v}_{k,1} \right) \geqslant f\left( \tilde{v}_{k,2} \right)}$ and, because $f(v)$ is monotonically increasing if ${v > 0}$, it follows ${\tilde{v}_{k,1} \geqslant \tilde{v}_{k,2}}$. The root $\tilde{v}_{k,2}$ is therefore the closest one to the origin.

We now show that $\tilde{v}_{k,2}$ decreases if $k$ increases by one. Because ${\frac{M-k}{k} > \frac{M-k-1}{k+1}}$, one has
\begin{equation}
	\label{eq:inequality_2}
	\frac{2 - f\left( \tilde{v}_{k,2} \right)}{\tilde{v}_{k,2}} > \frac{2 - f\left( \tilde{v}_{k+1,2} \right)}{\tilde{v}_{k+1,2}} \, , \nonumber
\end{equation}
i.e.
\begin{equation}
	\label{eq:inequality_3}
	\frac{2}{\tilde{v}_{k,2}} - \frac{2}{\tilde{v}_{k+1,2}} > \frac{f\left( \tilde{v}_{k,2} \right)}{\tilde{v}_{k,2}} - \frac{f\left( \tilde{v}_{k+1,2} \right)}{\tilde{v}_{k+1,2}} \, . \nonumber
\end{equation}
If it were ${\tilde{v}_{k,2} \geqslant \tilde{v}_{k+1,2}}$, the left-hand term of the previous inequality would be non-positive, whereas the right-hand term would be non-negative (the function ${f(v)/v = \ln\frac{1+v}{1-v}}$ is monotonically increasing if ${0<v<1}$), thus leading to a contradiction. It then must hold that ${\tilde{v}_{k,2} < \tilde{v}_{k+1,2}}$. Consequently, $\tilde{v}_{1,2}$ provides the minimum absolute value of a root of \Eq{eq:condition_v} and thus, via \Eq{eq:Lambda_0_new}, the maximum value of $\Lambda_0$ for a given $M$ with regard to the interior of the simplex $\mathbb{D}$.

As a corollary, the very same argument of the last paragraph can be promptly applied to show that increasing $M$ leads to a decrease of $\tilde{v}_{1,2}$ and thus to an increase of $\Lambda_{0,\,\text{max}}$. This corollary is important in order to discuss the behavior of the variance parameter $\Lambda_0$ on the boundary of the probability simplex.

It is finally important to note that, when ${k=1}$, the single \emph{outlier} among the set of probabilities ${\{s_i\}}$ is $s_M$, namely the one that was chosen as dependent on all the others [see \Eq{eq:p0}]. Because $M$ similar choices are possible, the number of equivalent maxima of $\Lambda_0$ is $M$.

\subsection{Boundary of the probability simplex}
\label{subsection:simplexBoundary}
The boundary ${\partial\mathbb{D}}$ of the $M$-simplex $\mathbb{D}$ is the union of $M$ boundary facets. Each facet is defined by a single $s_j$, with ${1 \leqslant j \leqslant M}$, set to zero, and it thus corresponds to a ${(M-1)}$-simplex.

This last assumption contradicts the fact that $M$ is the support size of the probability distribution. However, we discuss these cases for the sake of completeness and because, in principle, one could have arbitrarily small $s_i$ values.

Because each facet corresponds to a simplex, the arguments of the previous section can be iteratively applied as follows. First, in the interior of the $j$-th facet, the maximum of $\Lambda_0$ will be ${1/\tilde{v^\prime}_{1,2}^{2}-1}$, where $\tilde{v^\prime}_{1,2}$ is the negative root of \Eq{eq:condition_v} when ${k=1}$ and $M$ is replaced by ${M-1}$. Second, the boundary of the $j$-th facet has to be treated as a ${(M-2)}$-dimensional set of ${(M-2)}$-simplexes, and so on. The iteration stops at the two-dimensional simplexes in which only a pair ${(s_i,\,s_j)}$ of probabilities is nonzero: in this case the boundaries are the point-like vertices of a segment, in which ${\Lambda_0 = 0}$.

Because of the corollary expressed at the end of the previous section, Sec.~\ref{subsection:withinSimplex}, the maximum of $\Lambda_0$ on the $j$-th facet is strictly less than the value produced by a stationary point lying within $\mathbb{D}$, namely ${1/\tilde{v}_{1,2}^{2}-1}$.

\subsection{Maximum value of $\Lambda_0$ and related distributions}
\label{subsection:simplexComplete}
By taking into account the results of Sections~\ref{subsection:withinSimplex},~\ref{subsection:simplexBoundary}, we can conclude that the maximum of $\Lambda_0$ in the multinomial case occurs when the distribution of the probabilities ${\{s_i \,|\, i\in[1,\,M]\}}$ has a single entry equal to the value $p_0$ and all the other entries uniformly distributed and equal to $q_0$.

Upon finding the single positive root ${\tilde{v} \equiv \tilde{v}_{1,2}}$ of the transcendental equation
\begin{equation}
	\label{eq:condition_v_with_k_1}
	f(v) = 2 - v \ln (M-1) \, ,
\end{equation}
which corresponds to \Eq{eq:condition_v} with ${k=1}$, one has
\begin{eqnarray}
	p_0 &=& \frac{1 + \tilde{v}}{2} \, , \label{eq:p0_from_v_with_k_1} \\
	\Lambda_{0,\,\text{max}} &=& \frac{1}{\tilde{v}^2} - 1 \, . \label{eq:Lambda_0_new_with_k_1}
\end{eqnarray}

\begin{figure}[h!]
	\centering
	\includegraphics[width=0.45\textwidth]{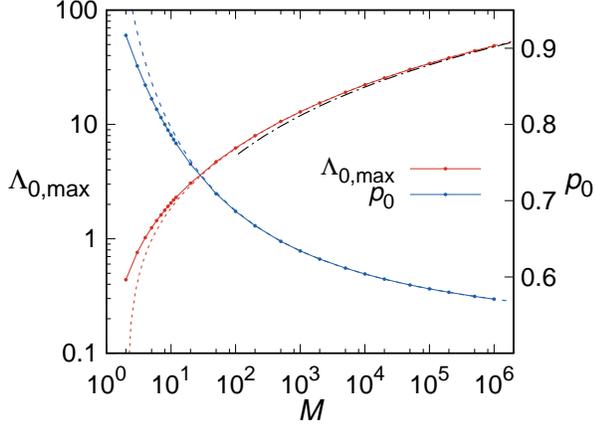}
	\caption{Maximum variance parameter $\Lambda_{0,\,\text{max}}$ as a function of the support size $M$: (red, solid line) exact computation; (red, dashed line) evaluation via \Eq{eq:Lambda_0_new_with_k_1} by using the approximated value for $\tilde{v}$ given by \Eq{eq:v_approximated_solution}; (black, dash-dotted line) approximation described by \Eq{eq:v_approximated_solution_2} and valid for ${M \gtrsim 100}$. \emph{Outlier} probability $p_0$ as a function of the support size $M$: (blue, solid line) exact computation; (blue, dashed line) evaluation via Eq.~(\ref{eq:p0_from_v_with_k_1}) by using the approximated value for $\tilde{v}$ given by \Eq{eq:v_approximated_solution}.}
	\label{fig:pAndLambdaZeroVsM}
\end{figure}

Equation~(\ref{eq:condition_v_with_k_1}) can be approximately solved by replacing $f(v)$ with the first term of its Taylor expansion, namely $2v^2$. Equation~(\ref{eq:condition_v_with_k_1}) then becomes a second-degree equation whose positive solution is given by
\begin{equation}
	\label{eq:v_approximated_solution}
	\tilde{v} \approx \sqrt{\frac{\ln^2(M-1)}{16} + 1} - \frac{\ln(M-1)}{4} \, ,
\end{equation}
which, for large $M$ values, can be further approximated as ${\tilde{v} \sim 2 / \ln(M)}$. (These approximations improve with $M$ because the larger $M$, the smaller $\tilde{v}$). Figure~\ref{fig:pAndLambdaZeroVsM} shows the plots of $p_0$, $\Lambda_{0,\,\text{max}}$ as a function of the dimension $M$. The approximation based on \Eq{eq:v_approximated_solution} works well for ${M \gtrsim 10}$.
It is worth observing that, as ${M\to\infty}$, one has
\begin{equation}
	\label{eq:v_approximated_solution_2}
	\Lambda_{0,\,\text{max}} \cong \frac{\ln^2(M)}{4} \, ,
\end{equation}
while $p_0$ tends to 1/2.

\begin{figure}[h!]
	\centering
	\includegraphics[scale=1.0]{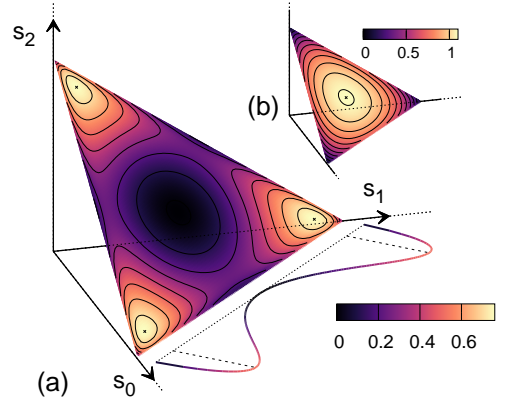}
	\caption{(a) Color map of the variance parameter $\Lambda_0$ on the probability simplex in the case ${M = 3}$. Three points, located at one coordinate being equal to ${p_0 \cong 0.88}$ and the other two equal to ${q_0 \cong 0.06}$, provide the maximum value of $\Lambda_0$, namely ${\Lambda_{0,\,\text{max}} \cong 0.762}$. The curve placed on the plane $\pi_{s_0,s_1}$ represents the plot of $\Lambda_0$ on the simplex facet corresponding to ${s_2=0}$. (b) Color map of Shannon entropy $H$ on the probability simplex in the case $M=3$. The maximum value occurs in the center, where ${s_0 = s_1 = s_2 = 1/3}$ and $\Lambda_0$ vanishes. No additional stationary points are present (which is true at any dimension).}
	\label{fig:varianceAndEntropySimplexes}
\end{figure}

The number of different distributions that maximize $\Lambda_0$ is equal to $M$, one for each state taking on the value $p_0$. Figure~\ref{fig:varianceAndEntropySimplexes} shows the case ${M=3}$.

It is important to note that, in the case of a distribution that maximizes $\Lambda_0$, the parameter $\Gamma$, which is given by \Eq{eq:Gamma} and describes the $N^{-1}$ dependence of the plug-in estimator $\widehat{\Lambda_0}$ of $\Lambda_0$, vanishes. A proof of this statement is discussed in the appendix. The situation is similar to what happens to the plug-in entropy estimator $\widehat{H}$ when the distribution is uniform: in this case $H$ takes on its maximum value equal to ${\ln(M)}$, while the quantity that describes the expected value of the variance of the plug-in estimator, namely $\Lambda_0/N$, vanishes and it is replaced by higher-order terms (typically ${\sim N^{-2}}$~\cite{Harris}). The same is expected to occur in the case of $\Gamma$.

Finally, it is worth considering how ``peaked'' the maxima of $\Lambda_0$ are. The partial derivatives with respect to the independent $s_i$'s, evaluated in the locations corresponding to ${\Lambda_0 = \Lambda_{0,\,\text{max}}}$ are given by
\begin{eqnarray}
	\label{eq:d2Lambda0_dpi_dpj}
	\frac{\partial^2 \Lambda_0}{\partial s_i \partial s_j} &=& - \left(\ln \frac{p_0}{q_0} \right) \left[\frac{\delta_{i,j}}{q_0} - \frac{1}{p_0} + 2 \left(\ln \frac{p_0}{q_0} \right)\right] \nonumber \\
	&=& - \Lambda_{0,\,\text{max}} \frac{4}{1 - \tilde{v}^2} \left[\delta_{i,j} (M - 1) \frac{\tilde{v}}{1 - \tilde{v}} + \frac{2 + \tilde{v}}{1 + \tilde{v}} \right] \, \nonumber \\
	&\approx& - 8 \Lambda_{0,\,\text{max}} \left[1 + \delta_{i,j} \frac{M}{\ln M} \right] \, , \nonumber
\end{eqnarray}
where the last approximation holds provided that ${M \gg 1}$ so that ${\tilde{v} \sim 2 / \ln(M)}$. In this limit, upon defining $\delta s_M$ and $\delta s_\parallel$ as
\begin{eqnarray}
	\label{eq:delta_s_0}
	\delta s_M &=& - \sum_{i=1}^{M-1} \delta s_i \, , \nonumber \\
	\label{eq:delta_s_parallel}
	\delta s_\parallel &=& \left[\sum_{i=1}^{M-1} \left( \delta s_i \right)^2 \right]^{1/2}\, , \nonumber
\end{eqnarray}
where $\delta s_i$ is the infinitesimal displacement of the respective $s_i$ from the value $q_0$, we have the following Taylor expansion for $\Lambda_0$:
\begin{equation}
	\label{eq:Lambda_0_TaylorExpansion}
	\frac{\Lambda_0}{\Lambda_{0,\,\text{max}}} \approx 1 - 4 \left[ \left(\delta s_M\right)^2 + \frac{M}{\ln M} \left(\delta s_\parallel\right)^2 \right] \, . \nonumber
\end{equation}

\section{Numerical experiments}
\label{section:numericalExperiments}
\begin{figure}[h!]
	\centering
	\includegraphics[width=0.45\textwidth]{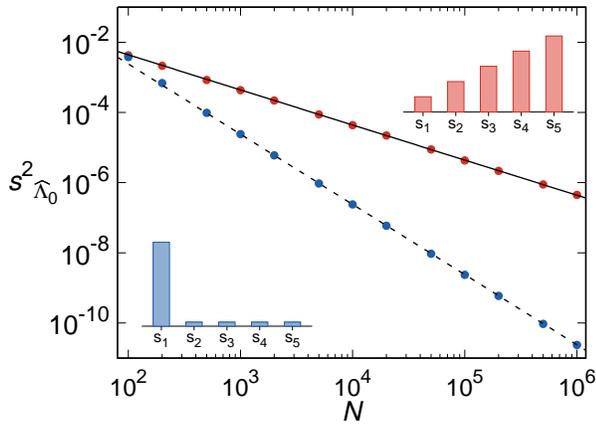}
	\caption{Variance of the plug-in estimator $\widehat{\Lambda_0}$ of the variance parameter $\Lambda_0$ as a function of the number of steps $N$ in the case of (red) the ``arithmetic progression'' distribution and (blue) the ``maximum variance'' distribution. Each dot corresponds to the sample variance of $\widehat{\Lambda_0}$ evaluated on a sample $\mathbb{S}$ of $10^4$ simulated evolutions of $N$ steps. The black, solid line correspond to the expected value of $s^2_{\widehat{\Lambda_0}}$ given by \Eq{eq:sampleVariance_plug-in_estimator}. In the case of the ``maximum variance'' distribution, the expected value of $\sigma_{\widehat{\Lambda_0}}^2$ vanishes (see main text). The black, dashed line corresponds to a $N^{-2}$ power law. Finally, the red (blue) inset shows a plot of the ``arithmetic progression'' (``maximum variance'') distribution.}
	\label{fig:varianceOfLambda0}
\end{figure}

To test the theory discussed in the previous two sections, two different distributions $\{s_i\}$ were considered, both with ${M = 5}$: an ``arithmetic progression'' distribution (red inset in Fig.~\ref{fig:varianceOfLambda0}), where $s_i$ is proportional to $i$; a ``maximum variance'' distribution (blue inset in Fig.~\ref{fig:varianceOfLambda0}), namely one of the distributions that yield a maximum $\Lambda_0$ in the case ${M = 5}$. Here, ${s_1 = p_0 \cong 0.834}$, whereas the other $s_i$'s are equal to ${q_0 \cong 0.042}$.

\begin{figure*}[t]
	\centering
	\includegraphics[width=0.9\textwidth]{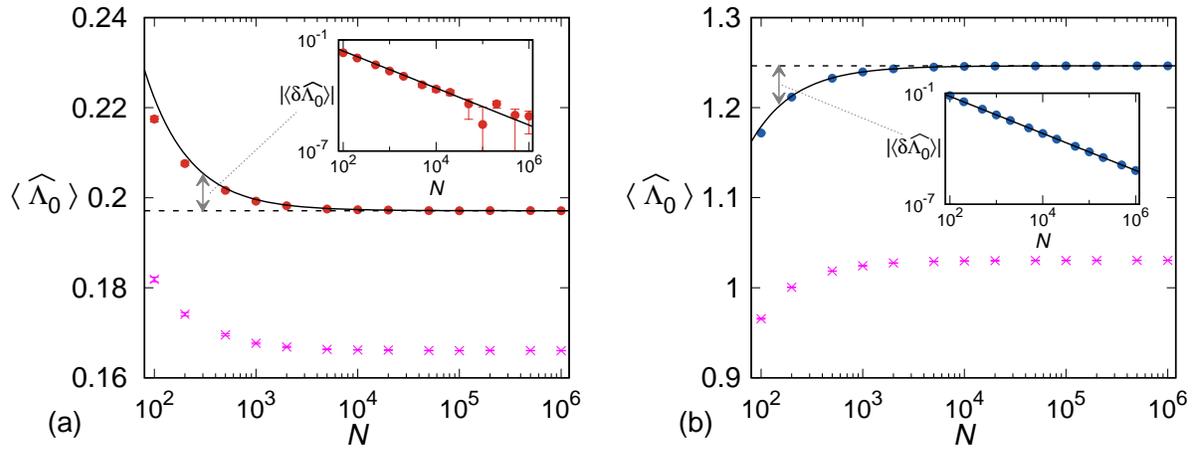}
	\caption{Plug-in estimator $\widehat{\Lambda_0}$ of the variance parameter $\Lambda_0$ as a function of the number of steps $N$ in the case of (a) the ``arithmetic progression'' distribution and (b) the ``maximum variance'' distribution. Each red [@(a)] or blue [@(b)] dot and the related errorbar corresponds to the sample mean and the sample standard deviation, respectively, of $\widehat{\Lambda_0}$ evaluated on a sample $\mathbb{S}$ of $10^4$ simulated evolutions of $N$ steps. The black, solid lines correspond to the expected value of ${\langle \widehat{\Lambda_0} \rangle}$ given by \Eq{eq:sampleMean_plug-in_estimator}. For both (a), (b), the insets show in log-log scale both the same numerical values and theoretical curves referred to the respective asymptotic theoretical values of $\Lambda_0$ [dashed lines: (a), $\Lambda_0 \cong 0.197$; (b), $\Lambda_0 \cong 1.246$]. Each magenta ``$\times$'' symbol and the related errorbar corresponds to the sample mean and the sample standard deviation, respectively, of the coefficient of $N^{-1}$ in Eq.~(40) of Ref.~\cite{Roulston} evaluated on the same sample $\mathbb{S}$.}
	\label{fig:biasOfLambda0}
\end{figure*}

For each of the two distributions and for values of $N$ ranging from $10^2$ to $10^6$, a sample $\mathbb{S}$ of $10^4$ evolutions were simulated, and for each evolution the plug-in estimator $\widehat{\Lambda_0}$ was evaluated. Thereupon the results were averaged on the sample $\mathbb{S}$ in order estimate the expected values of the mean and the variance of $\widehat{\Lambda_0}$, according to \Eq{eq:deltaEstimator_populationMean} and \Eq{eq:estimator_populationVariance}, respectively. By neglecting higher-order terms in those equations, the expected value of the sample mean ${\langle \widehat{\Lambda_0} \rangle}$ and the sample variance ${s^2_{\widehat{\Lambda_0}}}$ evaluated on a sample $\mathbb{S}$ are given by:
\begin{eqnarray}
\E\left( \langle \widehat{\Lambda_0} \rangle \right) \cong \Lambda_0 + \frac{\gamma}{N} \, , \label{eq:sampleMean_plug-in_estimator} \\
\E\left( s^2_{\widehat{\Lambda_0}} \right) \cong \frac{\Gamma}{N} \, . \label{eq:sampleVariance_plug-in_estimator}
\end{eqnarray}

Figure~\ref{fig:biasOfLambda0} and Fig.~\ref{fig:varianceOfLambda0} show the results of the numerical experiments for the mean and the variance of the plug-in estimator $\widehat{\Lambda_0}$, respectively. For both distributions in the case of the mean and for the ``arithmetic progression'' distribution in the case of the variance, the agreement with theoretical predictions is evident. In the case of the ``maximum variance'' distribution we have ${\Gamma = 0}$, so that the corresponding plot of $s^2_{\widehat{\Lambda_0}}$ in Fig.~\ref{fig:varianceOfLambda0} is expected to depend on higher-order terms in $N^{-1/2}$ than $N^{-1}$. Indeed we observe a $N^{-2}$ dependence.

In Fig.~\ref{fig:biasOfLambda0} the average values of $\widehat{\Lambda_0}$ are compared with the average values of the estimator proposed by Roulston and corresponding to the coefficient of $N^{-1}$ in Eq.~(40) of Ref.~\cite{Roulston}: ${\widehat{\Lambda}_{\text{Roulston}} = \sum_i \left[\ln \left(\widehat{p}_i\right) + \widehat{H}\right]^2 \widehat{p}_i (1-\widehat{p}_i)}$. This quantity was evaluated on the same data as $\widehat{\Lambda_0}$. The numerical results show that the estimator described in the present work is more reliable than Roulston's one in assessing the uncertainty of the plug-in estimator $\widehat{H}$ of Shannon entropy. The estimator proposed by Roulston indeed underestimates the variance of $\widehat{\Lambda_0}$: a reason is the fact that Roulston's expression of the variance is derived via propagation of error by assuming the observed number of visits to be mutually independent. However, as mentioned in Sec.~\ref{subsection:plugInEntropy}, the number of visits are constrained by the expression ${\sum_{i = 1}^M \widehat{j}_i = N}$, which leads to the set ${\{\widehat{j}_i\}}$ being distributed according to a multinomial distribution. The approach presented above takes into account this crucial property.

\section{Conclusions}
\label{section:conclusions}
The results obtained show that, under the null-hypothesis of an underlying multinomial distribution generating the data and provided that the size $M$ of the support is known, the expected uncertainty of an entropy estimation carried out via the plug-in estimator $\widehat{H}$ on a set of $N$ data, and neglecting higher-order terms in $N^{-1/2}$, is upper bounded as:
\begin{equation}
	\label{eq:uncertaintyUpperBound}
	\sigma_{\widehat{\Lambda_0}} \lesssim \left[ \frac{\Lambda_{0,\,\text{max}}(M)}{N} \right]^{1/2} \cong \frac{\ln(M)}{2\sqrt{N}} \, . \nonumber
\end{equation}
The upper bound can be then improved by using the plug-in estimator $\widehat{\Lambda_0}$ defined in \Eq{eq:sampleLambda0}. The evaluation of this estimator for different values of $N$ can deliver reliable estimates of the uncertainty affecting the assessments of Shannon entropy obtained by means of the plug-in estimator $\widehat{H}$.

Finally, it is worth noting that, in order to ``experimentally visit'' all the $M$ states, the number $N$ of trials has to exceed $M$, so that ${N > M > \sqrt{M}}$. Consequently, the upper bound described above turns out to be smaller, and thus more accurate, than the one derived by Antos and Kontoyiannis~\cite{AntosKontoyiannis}, namely ${(\ln N)/\sqrt{N}}$.

\appendix
\section{Proof of vanishing $\Gamma$ in the case of ${\Lambda_0 = \Lambda_{0,\,\text{max}}}$}
\label{section:appendix}
We show here that, in the case of ${\Lambda_0 = \Lambda_{0,\,\text{max}}}$, the parameter $\Gamma$ given by \Eq{eq:Gamma} vanishes. To this purpose we assume that, for a given $M$,
\begin{eqnarray}
	s_i &=& q_0 \quad\text{if}\quad 1 \leqslant i \leqslant M-1 \, , \label{eq:maximumVarianceDistribution_q0} \nonumber \\
	s_i &=& p_0 \quad\text{if}\quad i = M \, , \label{eq:maximumVarianceDistribution_p0} \nonumber
\end{eqnarray}
where, according to Eqs.~(\ref{eq:p0_bis}),(\ref{eq:solution_case2_bis}), it is:
\begin{eqnarray}
	p_0 + (M-1) q_0 &=& 1 \, , \label{eq:maximumVarianceDistribution_p0_q0} \\
	\ln p_0 + \ln q_0 + 2 + 2 H &=& 0 \, . \label{eq:maximumVarianceDistribution_logP0_logQ0}
\end{eqnarray}

Instead of relying on \Eq{eq:Gamma}, it is convenient to express $\Gamma$ as the coefficient of $t^2/2$ in \Eq{eq:exponentTaylorSeriesExpectedValue}:
\begin{equation}
	\label{eq:estimator_populationVariance_alternative}
	\Gamma = \sum_{i=1}^{M} \sum_{j=1}^{M} \chi_{i,j} \left[ \ell_i^2 + 2 (1 + H) \ell_i \right] \left[ \ell_j^2 + 2 (1 + H) \ell_j \right] \, . \nonumber
\end{equation}
By relying on \Eq{eq:maximumVarianceDistribution_logP0_logQ0} it is straightforward to show that each of the two terms ${\left[ \ell_i^2 + 2 (1 + H) \ell_i \right]}$, ${\left[ \ell_j^2 + 2 (1 + H) \ell_j \right]}$ appearing within the previous sum is equal to ${-(\ln p_0)(\ln q_0)}$. One then has:
\begin{equation}
	\label{eq:estimator_populationVariance_alternative_bis}
	\Gamma = (\ln p_0)^2 (\ln q_0)^2 \sum_{i=1}^{M} \sum_{j=1}^{M} \chi_{i,j} \, . \nonumber
\end{equation}
The sum, and thus $\Gamma$, is equal to zero due to the covariance matrix $\chi$ having vanishing column and row sums~\cite{Ricci2021CSF}. Alternatively, one can compute the sums by taking into account \Eq{eq:maximumVarianceDistribution_p0_q0}.

\bibliography{text}

\end{document}